

\input phyzzx
%
\baselineskip=16pt
\Pubnum={USITP-94-14 \cr}
\date={October 1994}
\titlepage
\title{The heavy quark potential in two-dimensional QCD with adjoint matter }
\bigskip
\author{T.H. Hansson\footnote\dag
{hansson@vana.physto.se; supported by the Swedish Natural Sience Research
Council }
and R. Tzani\footnote\ddag
{tzani@vana.physto.se; address after September 1994: Imperial College,
Physics Dept., Prince Consort Road, London SW72BZ}}
\address{Stockholm University, Department of Physics,\break
Box 6730, S-11385 Stockholm, SWEDEN}
\bigskip
\abstract{
Using a loop formulation approach of QCD$_2$,
we study the potential between two heavy quarks in the presence of adjoint
scalar fields, and demonstrate how 't Hooft's
planar rule is manifested in this
formulation. Based on some physical assumptions, we argue
that large adjoint loops ``confined'' inside an external fundamental one
give a Casimir type contribution to the potential energy, while the
small loops only renormalize the string tension. We also extend the results
to the case of massive adjoint fields. }

\endpage

\def\NP{{\it Nucl. Phys.\ }}
\def\PL{{\it Phys. Lett.\ }}
\def\PR{{\it Phys. Rev. \ }}
\def\PRD{{\it Phys. Rev. D\ }}

\def\PRL{{\it Phys. Rev. Lett.\ }}

\def\IJMP{{\it Int. Jour. Mod. Phys. A\ }}
\def\Mod{{\it Mod. Phys. Lett.\ }}

\REF\thoo{G. 't Hooft, {\it Nucl. Phys.}
{\bf B72} (1974) 461; {\bf B75} (1974) 461.}
\REF\thor{C.B. Thorn, \PL {\bf B99} (1981) 458.}
\REF\bars{I. Bars and A. Hansson, \PRD {\bf 13} (1976) 1744;
I. Bars, \PRL {\bf 36} (1976) 1521; I. Bars, \NP {\bf B111} (1976)
413.}
\REF\stro{A. Strominger, \PL {\bf 101B} (1981) 271.}
\REF\make{Yu.M. Makeenko and A.A. Migdal, \PL {\bf 88B} (1979) 135;
A.A. Migdal, \PR {\bf 175} (1980) 126.}
\REF\kaza{V.A. Kazakov and I.K. Kostov, \NP {\bf B176} (1980) 199.}
\REF\kazak{V.A. Kazakov, \NP {\bf B179} (1981) 283.}
\REF\polc{J. Polchinski, \PRL {\bf 68} (1992) 1267.}
\REF\gros{D. Gross, \NP {\bf B400} (1993) 161.}
\REF\mina{J.Minahan, \PRD {\bf 47} (1993) 3430.}
\REF\tayl{D. Gross and W. Taylor, \NP {\bf B400} (1993) 181 and \NP
{\bf B403} (1993) 395.}
\REF\poly{J. Minahan and A. P. Polychronakos, \PL {\bf B312} (1993) 155
and \NP {\bf B422} (1994) 172.}
\REF\doug{M. Douglas, RU-93-13, NSF-ITP-93-39, Mar. 1993. hep-th/9303159.}
\REF\gree{J. Greensite, \NP {\bf B249} (1985) 263 and \PRD {\bf 40} (1989)
4167.}
\REF\hans{T.H. Hansson and I. Zahed, \PL {\bf B309} (1993) 385.}
\REF\dall{S. Dalley and I. R. Klebanov, \PRD {\bf 47} (1993) 2517.}
\REF\deme{G. Bhanot, K. Demeterfi and I.R. Klebanov, \PRD {\bf 48} (1993) 4980
and \NP {\bf B418} (1994) 15.}
\REF\kuta{D. Kutasov, \NP {\bf B414} (1994) 33.}
\REF\migd{A. Migdal, Zh. Eksp. Teor. Fiz. {\bf 69} (1975) 810;
B. Rusakov, \Mod {\bf A5} (1990) 693.}
\REF\blau{M. Blau and G. Thomson, \IJMP {\bf A7} (1992) 3781.}
\REF\bard{K. Bardakci and S. Samuel, \PRD {\bf 18} (1978) 2849.}
\REF\hanss{A. M. Polyakov, {\it in Fields, strings and critical
phenomena,} Proc. Les Houches Summer School, Vol. IL 1988, ed. E.
Br\'ezin and J. Zinn-Justin (North-Holland, Amsterdam, 1990) p.305;
J. Grundberg, T.H. Hansson and A. Karlhede, \NP {\bf B347} (1990)
420.}
\REF\polya{A. M. Polyakov, \NP {\bf B268} (1986) 406.}
\REF\blot{H. W. J. Bl\"ote, J. L. Cardy and M. P. Nightingale, \PRL
{\bf 56} (1986) 742; I. Affleck, \PRL {\bf 56} (1986) 746.}
\REF\lusc{M. L\"uscher, \NP {\bf B180[FS2]} (1981) 317.}

\def\half{{1\over2}}
\def\a{\alpha}
\def\b{\beta}
\def\g{\gamma}

\def\t{\tau}

\def\w{\omega}
\def\X{\chi}
\def\noi{\noindent}
\def\ie{{\em i.e.} }
\noi
{\bf 1. Introduction}

Ever since the original work by 't Hooft [\thoo],  two-dimensional QCD (${\rm
QCD}_2$), has been an important ``laboratory'' for testing various ideas
about strong interaction physics. Early work established a string-like
meson spectrum and a parton picture for $e^+e^-$-production and
deep inelastic
scattering. The finite temperature behaviour was studied [\thor], and
the connection to string theory was worked out in some detail in a series
of papers by Bars and Hansson [\bars], who showed that to leading order in
$1/N$ 't Hooft's result follows from a string model with quarks attached to
the ends. Later, Strominger derived this string Lagrangian directly from QCD
by considering expectation values of Wilson loops and using factorization
at large $N$ [\stro].

In a related development Makeenko and Migdal [\make] reformulated Yang-Mills
theory in loop variables and Kazakov and Kostov [\kaza] managed to solve
the loop equations, for the case of QCD$_2$, and obtained closed equations
for the expectation values of arbitrary Wilson loops in the
 limit $ N \rightarrow \infty$. This approach was extended to finite
$N$ by Kazakov [\kazak].

Recently, there has again been a lot of interest in the connection between QCD
and string theory. In particular, by counting the degrees of freedom (d.o.f.)
in a QCD string, Polchinski has shown that a string theoretical description
of the large $N$ gauge theory would involve an infinite number of d.o.f.
at short distances [\polc].
Another line of research has focused on pure two-dimensional Yang-Mills
theory (YM$_2$). This theory does not have any field theoretical d.o.f.,
just as  strings in two dimensions do not have any transverse d.o.f..
On a compact space-time, however, the Wilson loops provide quantum
mechanical d.o.f., and it turns out that the corresponding partition
function, which can be calculated exactly, has a string
interpretation [\gros -\doug].

Yet another important development concerns two-dimensional Yang-Mills
theory coupled to scalar or fermionic {\em adjoint} matter (AdQCD$_2$).
AdQCD$_2$ is a very interesting theory from several points of view. First,
in the large $N$ limit, the planar adjoint loops are not suppressed, so the
planar approximation corresponds to a bona fide sum of fishnet
diagrams.\footnote\star{This should
be contrasted with QCD$_2$ where, in a linear gauge, the planar diagrams
include only dressed ladders that can be expressed by Dyson-Schwinger and
Bethe-Salpeter type equations with explicitly known kernels.}
To this day,
there is no known method to perform such sums analytically.\footnote\dagger{
For an interesting numerical approach in the content of a purely scalar
theory see [\gree].}

The second reason for studying (scalar) AdQCD$_2$ is that it is the high
temperature limit of pure YM$_3$, where the ``time''-component of the
original gauge field becomes an adjoint scalar in the dimensionally reduced
theory. As opposed to the dimensional reduction from QCD$_4$ to
scalar AdQCD$_3$, this reduction is not plagued by uncontrollable infrared
effects due to magnetic mass generation, but gives an exact description of
the static observables in the $T \rightarrow \infty$ limit [\hans].

In spite of not having any exact solution, even in the large $N$ limit, several
things are known about AdQCD$_2$:

\noindent
i) In  a series of papers, Dalley and Klebanov [\dall], and
Bhnot, Demeterfi and Klebanov [\deme] have numerically
studied AdQCD$_2$ in the large $N$ limit, using the light-cone
quantization technique of Brodsky and Pauli.
They obtained a discrete spectrum of bound states with a level density
vgrowing exponentially with the mass.

\noi ii)  Kutasov [\kuta], studied string properties of the theory
using the ideas of Polchinski [\polc] referred to above. He concluded that
the theory exhibits a deconfining phase analogous to Hagedorn transition
in string theory and that the spectrum consists of an infinite number of
Regge trajectories confirming, thus, its ``stringy'' nature.

The purpose of the present paper is to study AdQCD$_2$ using a  ``first
quantized''
formalism, where the partition function is a sum over loops describing the
paths of the adjoint particles. In this picture the gauge interaction
amounts to multiply each term in the sum with the expectation value of the
corresponding Wilson loops (in the adjoint representation). The reason for
taking this unconventional approach to an interacting field theory, is that
we will try to exploit the recently developed technology for computing
expectation values of Wilson loops. In fact, [\migd], [\kaza], [\kazak]
and [\blau] give methods for
calculating an arbitrary product of Wilson loops in an arbitrary
representation. The question is whether this can be put to practical use
in the study of  AdQCD$_2$.

Since our approach limits us to the study of
Wilson loops, the natural quantity to examine is the potential between heavy
particles. In particular, we shall study the heavy quark potential, both for
finite and infinite $N$. In the large $N$ limit, we know that the
string tension of QCD$_ 2$ with fundamental quarks is not renormalized by
quark loops (since they are down by
factors $1/N$), while for AdQCD$_2$ this is not the case.
In the ``first quantized'' version of
the theory, even this basic result is not obvious, but, as shown
in section 3, it follows from factorization of Wilson loop averages and
$N$ counting.  In section 4, we will use
our formalism to obtain  a non-trivial new result for the potential energy
between heavy quarks in the presence of an adjoint scalar field. For the
massless case it is:
$$
V(L) = \sigma_{eff} L -  {\pi \over {24}}  {1 \over L }
\eqn\pe$$
where $L$ is the distance between the quarks, and $\sigma_{eff}$ the
(non-calculable) effective string tension, which is renormalized by the
presense of adjoint loops.  We cannot give a strict
mathematical derivation of this formula, but it follows
from some rather reasonable assumptions. We want to stress that we do not
know of any way to derive this relation in the usual second quantized
formulation of the theory.

Before obtaining these results, we derive, in section 2, the
``first quantized'' version of the theory, starting from the more familiar
second quantized one. We also give some results for expectation values of
Wilson loops that will be used subsequently. For completeness we give
some details of how to derive these results in an appendix.
In section 5, we extend the
results to the case of massive adjoint field and
finally we summarize our results and discuss possible
extensions in section 6.

\bigskip\noi
{\bf 2. ``First quantized'' (loop) formulation of AdQCD$_2$}

We start from the partition function of 1+1-dimensional QCD coupled to
an adjoint scalar Higgs, $\Phi$, given by
$$
Z = \int {\cal D} A {\cal D} \Phi \; e ^ {- i S }
\eqn\pa
$$
where
$$
S = \int d ^2 x \, Tr \left[ {1 \over 4} F _ {ij}
F ^ {ij}
+ {1 \over 2} \left( \partial _ i \Phi - g [ A _ i , \Phi] \right) ^ 2
+ \half m ^ 2 {\Phi } ^ 2 \right]
\eqn\s
$$
and $A ^ \mu$ and $\Phi$ are in the adjoint representation of $SU(N)$,
$m$ is the mass of the Higgs field. We have chosen to study the theory
without any scalar self-interaction.
The $m$ appearing in the Lagrangian \s\ is a bare mass that cannot directly
be identified with the mass of the adjoint particles. We shall return to the
question of mass renormalization in section 4.
We shall be particularly interested in the case when
the renormalized mass vanishes, or at least satisfies $m \ll g$. It would be
interesting to study this limit both in perturbation theory and on the
lattice.

Note that in the dimensional reduction of our theory from a 3d
theory a mass term, a $\Phi^4$ term and higher order potential terms
$\sim\Phi^{2n}$ occur. The coefficients, which are functions of T, are
calculable in perturbation theory if proper care is taken to deal with the
infrared divergences due to the vanishing mass term.

The expectation value of an external Wilson loop $W_ {F}$ is  given by
$$
\left< W_{F} \right> _ {\rm QCD} = { 1 \over Z } \int {\cal D}  A{\cal D} \Phi
\, e ^ {i S} \, W _ {F}
\eqn\ev$$
where $S$ is defined by \s.

The next step is to convert the integration over $\Phi$ in
\pa\ to an integration
over loops. This is achieved by firstly integrating out the field $\Phi $
and then reexpressing the propagator in terms of particle variables as a
Feynman path-integral [\bard , \stro].
For the non-Abelian case with scalar particles we obtain
$$
\int{\cal D} \Phi \;  e ^ {- {i \over 2} \int d ^ 2 x
{\rm Tr} \left[  \left( \partial
_ i \Phi - g [ A _ i , \Phi]  \right) ^ 2 + \half m ^ 2 {\Phi } ^ 2
\right] } =\sum _ { n=0 } ^ {\infty} { 1 \over n ! }
$$
$$
\prod _ {l=1} ^ n
\int _ 0 ^ {\infty }  {dT _l \over T _ l} \int _{x_l(0)=x_l(T)}
D x _ l e ^ { i \sum _ {l=1} ^ n \oint _ 0 ^ {T_l} d \t _ l \left[ \half
m ^ 2 + \half \dot x _ l ^ 2 (\t _ l ) \right] }
Tr _ A \left(P e ^ { -i \oint d \t _ l A _ i ( x _ l)
\dot x ^ i _ l ( \t _l) } \right)
\eqn\ti$$
where $\t_l$  parametrizes the path, the dot stands for
derivative with respect to $\t_l$, the subscript $A$ means that the trace
is taken in the adjoint representation and $P$
denotes path-ordering. In obtaining the last expression
a regulator has been assumed and an infinite constant has been dropped.

Inserting \ti\ into \pa\ and rearranging the terms yields
$$
Z = \sum _ { n =0} ^ {\infty} { 1 \over n ! }
\prod _ {l=1} ^ n
\int _0 ^ \infty { d T _l \over T_l} \oint
D x _ l \; e ^ { i \sum _ {l=1} ^ n \oint _0^ {T_l} d \t _ l \, [ \half m ^2
+ \half \dot x _ l ^ 2 (\t _ l) ] } \,
\left< \prod _ {l=1} ^ n W _ {A_l} \right>
\eqn\pe
$$
where the last expectation value
of Wilson loops is with respect to the pure Yang-Mills action, and the
normalization is such that $\left< 1 \right>  =1 $.

Finally, using \ev\ and \pe, we can derive the following expression for
the expectation value $\left< W_F \right>$ of a Wilson loop:
$$
\left< W_{F} \right> _ {\rm QCD} = \sum _ { n =0}
^ {\infty} { 1 \over n ! }
\prod _ {l=1} ^ n
\int _0 ^ \infty { d T _l \over T_l} \oint
D x _ l \; e ^ { i \sum _ {l=1} ^ n \oint _0^ {T_l}
d \t _ l \, [ \half m ^2
+ \half \dot x _ l ^ 2 (\t _ l) ] } \,
\left< W_F \prod _ {l=1} ^ n W _ {A_l} \right>^{c}
\eqn\ev
$$
where the connected part of the expectation value of products of Wilson
loops is defined as connected with respect to the external loop, \ie
$$
\eqalign{ \left< W_F W_A \right> ^c & = \left<W_F W_A \right> -
\left< W_F \right> \left< W_A \right> \, , \cr
\left< W_F W_{A_1} W_{A_2} \right> ^c
& =  \left<W_F W_{A_1} W_{A_2} \right>
-\left< W_F W_{A_1} \right>^c\left<  W_{A_2} \right>
-\left< W_F W_{A_2} \right>^c\left<  W_{A_1} \right>  \cr
& -  \left< W_F \right>   \left<W_{A_1} W_{A_2} \right> \, , \cr
 &{\rm etc}. \cr}
\eqn\cd
$$
The fact that the expectation value \ev\ depends only on the connected
loops can be shown by standard arguments. Similarly
the partition function can be expressed as,
$$
Z= e ^ { \left< W \right> + \half \left<WW\right> ^ c + {1 \over 3 !}
\left<WWW \right> ^ c + ... } \ \ \ \ \ ,
\eqn\ff$$
where the superscript $c$ here stands for the fully connected averages.
$$
\eqalign{\left< WW \right> ^c &\equiv \left<WW \right> -
\left< W \right> \left< W \right> \, , \cr
\left< WWW \right> ^c
&\equiv \left<W W W \right>
-3 \left<W \right> \left< W W  \right>
+ 2\left< W \right> \left< W \right> \left< W \right> \, , {\rm etc}. \cr}
\eqn\con
$$
The formula \ff\ is essentially an expansion in terms of cummulants.

\noindent
Comments:

\noindent
The relations \pe\ and \ev\ define a massive ${\rm AdQCD}
 _ 2$ in terms of
classical trajectories of particles of mass $m$ and Wilson loop averages
in the adjoint representation.

\noi
The expression for $\left< W_F \right> _ {\rm QCD}$ is in
terms of connected green functions, just as an S-matrix element in ordinary
field theory.

\noindent
For the case of spinning particles the integrand in the right hand side
of \ti\ must be multiplied by a
spin factor. In two dimensions this is simply $ e^{i \pi (\nu +1)}$,
where $\nu$ is the number of self-intersections of the loop, as
originally obtained by Strominger [\stro]. For a general discussion
of spin factors in higher dimensions, see \eg  [\hanss].

\vskip 3mm \noindent
{\bf 3. Expectation values of Wilson
loops and the $N\rightarrow \infty$ limit}

As mentioned in the introduction, we will
try to gain a partial understanding of the theory defined by \pe .
For this purpose, we shall now discuss expectation values of Wilson
loops of the type occuring in \pe\ and \ev.
These can be obtained using methods given in [\migd , \kaza , \poly ,
\doug , \blau] and we shall here only state the results.
For completeness, some of the techniques and sample
calculations are summarized in the appendix.

We first consider the configuration of $n$ disconnected adjoint loops
inside the external fundamental one, see fig. 1a.
We can assume the space outside
of the external loop to have arbitrary topology, but we take its area
to be infinite. Following the method discussed in
appendix we derive the expression:
$$
\eqalign{ &\left< W_F W_{A_1} ....W_{A_n} \right> = \cr
& { {e ^ {- {g_0^2 \over N} (S_F - \sum_{i=1}^n S_i) C_F } }
\over { (d _F)^{n-1} } }
\prod_{i=1}^n
 \left( d _ F e ^ { - {g_0 ^2 \over N}  S _i C _F}
+ d _ {R_1} e^{ - { g_0 ^ 2 \over N }  S _i C_{R_1} }
+ d _ {R_2} e^{ - { g_0 ^ 2 \over N } S_i C_{R_2}} \right) \cr }
\eqn\am
$$
where $S_i $ is the area inside the $i^{th}$  adjoint loop and $S_F$
the area of
the fundamental loop. By $C_R $ we denote the quadratic casimir
of the representation $R$ and $R_1$, $R_2$ are defined by the
decomposition $F\otimes A = F \oplus R_1 \oplus R_2$.
We have also redefined
the ${\rm QCD}$ coupling constant $g^2$ by ${g_0 ^2 \over N}$, so that
in the large $N$ limit $g _0^2 $ remains finite.

Before we proceed to  present more results of a similar kind, it is
instructive to give a string interpretation of the relation
\am.
Remember that the expectation value of a Wilson loop in the fundamental
representation with respect to pure Yang-Mill's action is
$
\left< W _ F \right> = d _ F e ^ {- {  g _0 ^ 2  \over N } S_F C_F }
$
so the  tension of a fundamental string is,
$\sigma _F = {g_0^2 \over N } C _F$, which in the large $N$ limit
equals  ${g_0 ^2 \over 2}$.
Similarly the exponents of the three terms inside the parenthesis
in \am\ can be associated with the three
different values for the string tension corresponding to the three
different representations in the product $F \otimes A$. The corresponding
prefactors $d_R$ give the probability for the different tensions. In the
large $N$ limit where an adjoint particle can be thought simply of as a
$\overline F F$ combination, we have  the string configurations
$\overline F-A-A-F$ and $\overline F-A\equiv A-F$, corresponding to
$\sigma_F$ and $ \sigma_ {R_1} = \sigma_ {R_2} =
3 \sigma_ {F} $, respectively.

We also give the result for the
configuration with two non-intersecting adjoint loops, one inside the other,
see fig. 1b. It is given by
$$
\eqalign{\left< W _F W _ A W _ A \right> \,&=\, \cr
e ^ {-{g_0^2 \over N} (S_F-S_2) C_F } &\left( e ^ {-{g_0^2 \over N}
(S_2-S_1) C_F}
\left[d_{R_1} e ^ {- {g_0^2\over N} S_1 C _{R_1}} + d_{R _2} e ^ {-
{g_0^2 \over N} S_1 C_{R_2} } + d_F e ^ {-{g_0^2 \over N } S_1 C_F } \right]
\right. \cr
+
\left. e ^ {-{g_0^2 \over N} (S_2-S_1) C _ {R_1}} \right.
& \left. \sum _{K_i} d_{K_i} e ^ {-
 {g_0^2 \over N}
S_1 C _ {K_i}} + e ^ {-{g_0^2 \over N} (S_2-S_1) C _ {R_2} }
\sum _ {L_j} d_{L_ j}e ^ {-{g_0^2 \over N } S_1 C _ {L_j} } \right) \cr }
\eqn\aa$$
where $S_1$ and $S_2$ are the areas of the adjoint loops (with
$S_2 >S_1$),
$S_ F$ is the area of the fundamental, $K_i$ and $L_j$ are
the irreducible components obtained from the decompositions of $ R _1
\otimes A$ and $R _2 \otimes A$ respectively.
The generalization of this to n adjoint loops with
one inside the other is straightforward, but complicated to write
in a closed form. We see that as one adds more and more
adjoint loops, one inside
the other, the string tension receives contributions from higher and higher
representations. All contributions are multiplied by an entropy factor,
which is simply the dimension of the representation.

Expectation values for (self)intersecting loops are harder to calculate,
and the methods are briefly discussed in the appendix. We again record
some results. For the configuration  of two  intersecting adjoint loops
of areas $ S_1 +S $ and $ S_2 +S$, see fig. 2a, we get
$
\left< W _A W _A \right> = e ^{-{g_0^2 \over N } S _1 C_A }
\, e ^ {-{g_0^2 \over N }S _2 C_A } \,
\sum _ i d _{A _i} e ^ {-{g_0^2 \over N } S C _ {A_i} }
$,
where $A_i$ are defined by $ A \otimes A = \sum _i A_i $.
For a self-intersecting loop of total area $S_1 + S_2$ and
one point of intersection, as shown in fig. 2b, it is
$
\left< W _A \right> = d_ A \, e ^ {-{g_0^2 \over N} (S_1+S _2) C_A}
$.
This last result can be trivially generalized
to a loop with $n$ self-intersectings but no overlapping areas.
In this case the area in the exponent of the last expression
is replaced by the total area ($ \sum _{i=1} ^{i=n} S_i $).
In fig. 2c, finally we show a self-intersecting loop with overlapping
areas. It is
$
\left< W _A \right> = \, e ^ {- {g_0^2 \over N } S_1 C_A } \sum _ {R_i}
d _{R_i} \a _ i \, e ^ { - {g _ 0^2 \over N } S_2 C _ {R_i} }  \,
$,
where $R _i$ and $\a _ i $ are defined by the decomposition
$ Tr _ A (U ^2) = \sum _ {R_i} Tr _ {R_i} (U) \a _i $.

For large $N$, expectation values of products of Wilson loops
factorize. This follows from general arguments and can easily be
shown in our formulation, using some simple group
theoretical rules given in the appendix. For instance,
the relations \am\ and \aa\ become
$$
\eqalign{
\left<W_F W_A ...W _A \right>  &= d _F e ^ {-{g_0^2 \over N} S_F
C_F } \prod _ i d _A e ^ { - {g_0^2 \over N } S _i C_A }
\left[1 + O(1/N)\right] \cr
\left< W_F W _A W _A \right>  &= d _ F e ^ {-{g_0^2 \over N} S_F
 C _F} d _ A e ^ { - {g_0^2 \over N }  S _2 C _A }
d _ A e ^ {-{g_0^2 \over N } S_1 C _A } \left[1 + O(1/N)\right]
\cr }
\eqn\ns
$$
Due to this factorization, the leading term in $N$ will
cancel in the relations \cd\ .
The sub-leading contributions can however
not be ignored in the calculation of $ \left< W _ F
\right> _ {QCD}$, since they are
of the same order as $\left<W_F \right>$.
To see this,  notice that in $ \left<
W _ F W _ A \right> ^ c $ it is the sub-leading part
of $ \left< W _ F W _ A \right> $ which contributes, in
$\left< W _ F W _ A W _ A \right> ^ c $ it is the sub-sub-leading
term of $ \left< W_F W _ A W _ A \right> $ which contributes and so on.
The sub-leading part of $ \left< W_F W _A \right> $, the sub-sub-leading
part of $\left<W _F W _ A W _ A \right> $ and so on are $O(N)$, thus
all these connected expectation values are $O(N)$.
Expressions similar
to \ev\ and \ns\ can easily be found for the 't Hooft model,
\ie for insertion of fundamental loops. The difference in this case is
that the sub-leading term is $O(1)$, and can thus be neglected in the
large $N$ limit in the
calculation of the expectation value $ \left< W _ F \right> _ {QCD}$.
We have therefore obtained, in the loop formulation, the well
known result that the fundamental loops do not renormalize the string tension,
while the adjoint ones do. Similarly, the sub-leading parts of the
connected averages in \ff\ can be neglected for the case of
fundamental loops but cannot be neglected when the
loops are in the adjoint representation.
This, again, is a manifestation of the 't Hooft
planar rule in the loop formulation.

\bigskip\noi
{\bf 4. Large loop contributions to $W_F$ }

In this section we shall estimate the contribution from a certain
class of large  (compared to $ g^{-2} $)
adjoint loops to the expectation
value of a large fundamental loop, and thus to the static potential between
heavy ``quarks''. We start with the following observations:

\noi$ i)$
In the expectation value $\langle W_F W_{A _1}.....W_{A_n}\rangle$, where all
the adjoint loops are contained within the fundamental loop,
there is always {\em one} term
$\sim e^{-g^2 S_F C_F}$ since $F$ occurs {\it once} in the decomposition
$F\otimes A$.
In the string language, this corresponds to the
possibility of putting an adjoint particle somewhere in a fundamental
string without changing its tension on either side.

\noi$ii)$
For expectation values $\langle W_F W ^ L_{A_1}...W ^ L _{A_n}\rangle$
where all adjoint loops are large, \ie
with an area $\gg g^{-2}$, the above term will be exponentially large compared
with all other contributions.
A simple example is shown in fig. 1a.

\noi$ iii)$
Configurations with at least one loop extending outside the
fundamental loop are exponentially suppressed as illustrated in
fig. 3.

\noi
Even though true for the configurations in fig. 1a and 2b, the statements
{\em ii)} and {\em iii)} are not strictly true. We will return to this
question below.

\noi
The  basic idea in this paper is as follows:

Consider a configuration with only large loops, $W^L_A$,
that are all confined within the fundamental one.
For this case we can, with exponential accuracy, make the
replacement
$$
\langle W_F W_{A_1}^L.....W_{A_n}^L \rangle  \rightarrow
e^{-g^2 C_F S_F} \, ,
\eqn\sk$$
at least in the generic case where the areas of
intersection are also large. In the presence of small loops that do not
intersect the large ones, we similarly have
$$
\langle W_F W_{A_1}^S.....W_{A_m}^S  W_{A_1}^L.....W_{A_n}^L \rangle
\rightarrow  \langle W_F W_{A_1}^S.....W_{A_m}^S \rangle \, .
\eqn\ska$$
Let us for the moment neglect the small loops that intersect the large ones
and try to understand the meaning of \sk\ and \ska.
That we can remove the large loops from the gauge field averages
means that they  are non-interacting. The string version of
this statement is that an adjoint particle can be put on the string of
fundamentals in such
a way that it feels no net force, since the string tension is the same on
both its sides. This does not mean, however, that the large loops
correspond to those of a
completely free field theory since they have to be ``confined'' withing the
contour of the fundamental loop in order not to be suppressed. Below we
shall argue that the sum of these non-interacting, but ``confined loops'',
is simply related to the Casimir energy of a free field theory.

Let us now return to the small loops intersecting one or several of the
large ones. Those intersecting a single large loop will generically
give a contribution that is proportional to the length of the large
loop. It is natural to identify this as a mass
renormalization of the particles in the large loops. Similarly the small
loops intersecting two large loops will give rise to a short range interaction
between them. Neither of these statements can be proven, but they
are reasonable assumptions. In the following we will be interested in the
long wavelength properties of the theory, and therefore keep the mass
renormalization but  neglect any short range  interaction. We
will comment on this approximation in the last section.

In order to express these assumptions formally, we
separate the integration over $x_l$ in the exponent of the
expression \ev\ into large and small loop contribution. Then,
according to the discussion above, the Wilson loops $W_{A_i}^L$ for the large
loops can be removed from  the expectation value in the presence of the
external fundamental loop. (Notice that with an exponential accuracy
the large Wilson loops can be also removed from the connected expectation
values.) Thus \ev\ can be written as
$$
\left< W_F\right> \simeq e ^ {\int_0 ^ \infty {dT \over T}
\int_{Large} D x
e ^ {i S_{r} } } \left< exp \left( \int _0 ^ {\infty} {dT \over T}
\int _{Small} D x e ^ {i S_ {b}} W _A \right) W _ F \right>^c
\eqn\ga$$
where the subindices $r$ and $b$ in the definition of $S_r$ and
$S_b$ refer to the renormalized and bare mass respectively; that is,
$ S_r = \int _0 ^T d\t (\half m^2 _r + \half \dot x ^2 ) $
and $ S_b = \int _0 ^T d\t (\half m^2 _b + \half \dot x ^2 ) $.
Let us emphasize again that the integral in \ga\ is taken over
loops contained inside the fundamental loop.

We now make the following assumptions:

\noi
$i)$ The large loops that are ``confined'' within the contour of the
fundamental loop can effectively be taken into account by imposing
appropriate boundary conditions. It will be important that the quantities we
calculate do not depend on the details of these conditions, as will be
discussed below.
\medskip\noi
$ii)$ The (non-calculable) last factor in  \ga\ will effectively only
renormalize the string tension. This is almost trivially true, since the
tension is the only quantity that characterizes the low-energy behaviour of
a one-dimensional string. (In higher dimensions there could be terms related
to extrinsic curvature [\polya], and for highly excited states one will of
course ``see'' the adjoint degrees of freedom.)

Since the expectation value  in \ga\ that
depends on the small loops factorizes, and has the form of the
Wilson loop average (7),  assumption ii) implies that
it will contribute a factor  $e^{-\sigma(R) S_F}$ to $\left<W_F\right>$.
Here the
string tension $\sigma(R)$ depends on the cutoff $R$
which defines the meaning of large and small.
{}From the previous discussion it should be clear that we must regulate
in such a way as $ L >> R >> { 1 \over g } $.

The other factor in \ga, that depends on the large loops, has the form of a
partition function of a confined scalar field theory, and is thus directly
connected to the vacuum energy which, as we shall argue below, is of the
form
$$
E_{vac; 2} = \, { \lambda \over R^2} \, L   - \, {\pi \over 24} \, {1 \over
L}\, ,
\eqn\casx
$$
where $\lambda$ is a constant and the subindex 2 stands for the dimension of
space-time.
Thus, assuming i) and ii), we can write an
explicit formula for a rectangular Wilson loop with area $S=LT$ in the
limit $T\gg L \gg g^{-1}$:
$$
\left< W_F\right> = e^{-E_{tot} T } \, ,
\eqn\res
$$
where
$$
E_{tot} = \sigma_{eff} L + 2 \delta M - {\pi \over {24}}  {1 \over L }
\eqn\etot
$$
is the total energy of a color singlet pair of static quarks.
The last term in \etot\ is the Casimir energy of a
free, massless field confined to
the line-segment $L$ and $\delta M$ is a non-calculable mass
renormalization of the heavy external source. This will arise from small
loops intersecting the external in analogy with the mass renormalization of
the large adjoint loops. Note that the term $\sim L/R^2$ in \casx\ must
cancel against an identical piece in the cutoff dependent term
$\sigma(R)$L coming from the small loops, in order to give a cutoff
independent effective string tension $\sigma_{eff}$.
Finally, notice also that the approximation of the large loops being free
holds only for finite $N$. All the results in this section are thus true
only for finite $N$.

We now explain how we arrived at \casx.
The vacuum energy of a free field theory
defined in a  spatial volume $V$ is ultraviolet divergent.
In addition to the usual free space divergence
 $\sim V \Lambda^d$, where $\Lambda$ is an
ultraviolet cutoff and  $d$ the space-time dimension, there are less
singular terms depending on the particular boundary.
In our case, \ie for $d=2$, the expansion is
$$
E_{vac; 2} = L \Lambda^2   + {c \over L }
\eqn\cas
$$
A convenient way to exhibit the  divergence structure is by using the
so called multiple reflection expansion (MRE) method.
The vacuum energy is related
to the Euclidean propagator $ G ( x, x ^ {\prime } ; i\beta ) $, with
$\b$ denoting the Euclidean time, via,
$$
E_{vac} = lim _ {\beta \rightarrow 0} { 1\over 2} \int d x
_ {x ^\prime  \rightarrow x }
(- \partial _ {\beta} ^ 2 + \nabla _x ^ 2 ) G (x,x ^ {\prime} ; i\beta )
\eqn\va
$$
where $G(x, x ^ {\prime} ; i\beta)$ is defined with the appropriate boundary
conditions at the end points of the segment.
In the MRE-method one expands this propagator in a
series in n, where the $n ^ {\rm th}$ term is the configuration with
n reflections in the boundaries and free
propagation between the reflections.
For a scalar field in $1+1$ dimension, the propagator is given by
$$
G  ^ N (x , x^ \prime ; i\beta) = \sum _ {n= - \infty} ^ {n= \infty}
(\pm1)^n G ^ 0 ( x, x ^ \prime _ n ; i\b)
\eqn\mp$$
where $ x ^\prime _ n = (-1) ^ n x ^\prime + n L$, with
$|x'| < { L \over 2}$, is the position of the $n$th mirror image
of the point $x ^ \prime$ and $L$ is the volume. The $+$ and $-$ signs
refer to Neuman and Dirichlet boundary conditions respectively.
It is then straightforward to show that the leading divergence in \cas\
comes from the $n=0$ term in \mp\
and the Casimir energy from the terms with $n$ even.
(The terms in the expansion with $n$ odd give zero.)
Summing the contributions with at least two reflections, we get
$$
E_{vac} = {2 \over \pi \beta ^ 2 } L - { \pi \over 24 } \, {1 \over L}
\eqn\aaa
$$
irrespectively of whether the boundary conditions are Dirichlet or Neuman.
The origin of the last term in \etot\ should now be clear.
It is the Casimir energy.
It is an assumption that the ``confinement'' of the loops, which in the
QCD case is due to the exponential suppression of loops extending beyond
$W_F$, can be replaced by a simple, ``confining'' boundary condition.
In the framework of the MRE this is very natural since we can think of the
exponential supression as providing a reflection of the particle, and
summing over all loops will, as usual, reproduce the quantum mechanical
propagator. In this
context it is pleasing that, in the two cases we have considered, the result
does not depend on the specific boundary condition. For practical
calculations one uses either a momentum cutoff or a point splitting
procedure. However, since the divergence structure cannot depend on the
details of the cutoff, a cutoff in the length of the loops should be
equally good, so we can replace $\Lambda$ with $1/R$.
Notice that our result for the Casimir energy is consistent with the
finite scaling results of quantum field theory [\blot]. In fact,
every conformal field theory in $1+1$
dimensions with central charge equal to $1$ contributes a factor equal
to $- { \pi \over 24 L}$ as
the finite-size corrections to the free-energy of an infinite long
strip of width $L$, when the boundary conditions are appropriately
chosen.

\bigskip\noi
{\bf 5. The massive case}

So far we have considered only the $m_r =0$ case. The above analysis,
however, can be extended for the case where $m_r \neq 0 $. As we shall
see, the most important effect, i.e. the $L$-dependent part in the
Casimir energy in \etot, will be negligible in the limit where $ L >> m _ r
^ {-1} $.
The vacuum energy for the case of massive scalar field
is given by
$$
E _{vac}= \sum _ n \half \, \w _ n  ={\rm  lim} _ {\beta \rightarrow 0 }
( - { \partial Z \over \partial \beta } )
\eqn\cd$$
where $ Z = \sum _ n e ^ {- \beta { \w _ n \over 2} } $ and
$ \w _ n = \sqrt { k _ n ^2 + m ^ 2 },  $ with $ k _n = { \pi n \over
L } , n = 1, 2,...$.

{}From \cd\ we obtain
$$
\eqalign{
E _ {vac} &=  { 2 \over \pi \beta ^ 2 } \, L  - { m ^ 2 \over 4 \pi }
\ln ( {m \beta \over 4 })\, L - { 3 m^ 2 \over 8 \pi } \, L + { \g m ^ 2
\over 4 \pi } \, L
+{ \pi \over 48 L } ( 1 + { 4 m^2 L ^2 \over {\pi} ^2 }) ^
{-\half}  \cr
&- {\pi \over 16 L } \sqrt { 1 + {4 m ^2 L ^ 2 \over {\pi} ^ 2 }}
- { m ^ 2 L \over 4 \pi } \ln \left[ { \pi \over 2 m L } ( 1 +
\sqrt { 1 + {4 m ^ 2 L ^ 2  \over {\pi} ^ 2 } } ) \right] \cr }
\eqn\mc$$

To get some feeling for  this result we shall consider different limis.
First, when $mL$ goes to zero, with $L$ fixed, we get
$$
E_{vac} \rightarrow  { 2 \over \pi \b ^ 2 }\, L
- { m ^ 2 \over \pi } \ln ({\b m \over 2})\, L + \left(
- {5 m^2 \over 3 \pi} + { \g m ^2 \over \pi ^2 } + { m^2 \over
4 \pi } \ln ( {mL \over \pi}) \right) L
- {\pi \over 24 } { 1 \over L}
\eqn\sm
$$
In the limit $ m \rightarrow 0$ this gives the massless result \aaa,
as expected. Notice that
the terms proportional to $L$ in \sm\ contribute to the string tension
renormalization, while the last term can be interpreted as the
Casimir energy of a massive field.

In the limit where $mL >> 1$ the expression \mc\ becomes
$$
E_{vac} \rightarrow  { 2 L \over \pi \b ^2 } - \left(
 { 1 \over 4 \pi } \ln ({m\b \over 4 })  + {3 \over 8 \pi}
+ { \g \over 4 \pi } \right) m ^ 2 L - { m \over 4}
\eqn\la$$
In order for the last limit to have a meaning, however, we must have
$ m << {1 \over \b}$,
since here $ \b$ plays the role of $R$, the cutoff for the size of the
loops, and $L \leq  {1 \over m} $, which is the Compton wave length.

As in the massless case we take for granted that the cutoff dependent
pieces in \mc\ combines with the non-calculable contributions from the small
loops to give a cutoff independent effective string tension. Thus we have a
prediction for the Wilson loop average for any value of the renormalized
mass.

We conclude this section with a calculation of the effective string tension
in the limit where $ m >> g $.\footnote*{
This calculation was suggested to us by Ismail Zahed.}
For this purpose we
expand the effective action of the gauge field, $ \Gamma [A] $
in terms of $ {g ^2 \over m ^2}$. The leading contribution comes from
the one-loop, vacuum polarization tensor, which is
$ \sim {g ^ 2 \over m ^2} $; namely, $ \Pi ^ {\mu \nu} (k) = ( g ^ {\mu \nu
} k ^2 - k ^ \mu k ^ \nu ) \Pi (k ^2) $.
In this case, the expansion gives
$$
\Pi (k ^2) =  - {g ^2 \over m ^2} {1 \over 12 \pi } \left( g ^ {\mu \nu}
k ^2 - k ^ \mu k ^ \nu \right) \, + \, { g ^2 \over 2 \pi} g ^ {\mu \nu }
\left( {\rm ln} {4 \pi \mu ^2 \over m ^2} + \Psi (1) \right)
\eqn\pv$$
Therefore the renormalized propagator is given by
$$
{ g ^ 2 \over k ^ 2 } \, { 1 \over 1 + \Pi (k ^ 2)} \, \simeq \, { g ^2 _ {eff}
\over k ^2 }
\eqn\rp$$
where
$$
g ^ 2 _ {eff} = g ^ 2 (1 + { C _ A \over 12 \pi } { g ^ 2 \over m ^ 2}) \,
\eqn\rs$$
and $C _A$ is the second Casimir of the adjoint representation.
Since the correct Coulomb energy is $ E _C \sim g ^ 2 _ {eff} L $, it is
clear, that the expression \rs\ gives a correction to the string tension.
Note that the effective string tension increases due to the presence of the
heavy scalar.

\bigskip\noi
{\bf 6. Discussion and outlook }

It should be clear from the previous two sections that our results for
the heavy quark potential rests on heuristic arguments, and we want to
comment upon their character.

First, we assumed results derived for specific
configurations of Wilson loops to
hold even in cases where we have not done explicit calculations. In
particular, we assumed that curled up and intersecting loops
will not qualitatively change the picture derived from configurations
of smooth and non-intersecting ones. Such an assumption could well be
dangerous, since we have no way of controlling the number of
configurations of different types. Put differently, our arguments are all
based on ``energy'' while we have no possibility to even estimate
``entropy'' effects. At least in a continuum
theory, we know of no loop space techniques that would allow
us to attack this problem - not even in principle.

Second, we made certain assumptions about the effective large-distance
theory of ``large'' loops. Some of these, like the
presence of an effective string-tension and a mass renormalization of the
large adjoint loops, we think that are rather safe.
Our way to treat a sum of large confined and
non-interacting loops by calculating vacuum energies \'a la Casimir
is supported by the multiple reflection type arguments given in the text.
However, the assumption that the large loops are non-interacting, on which
this calculation rests, can certainly be doubted. It would be interesting
to check, using perturbation theory, if our results are stable against
adding a weak short range interaction in the vacuum energy calculation.
However, calculations with both boundaries and interactions are
notoriously complicated and, even in two dimensions, it is a non-trivial
task.

Having  given all the warnings, we shall now stress some interesting
aspects of our result.

The presence of a  $\sim L^{-1}$ type correction to the string tension
certainly is very reminiscent of L{\"u}schers' universal finite size
correction term in string theory [\lusc]. In this context the correction is due
to
the transverse oscillations of the string, but the calculation is again
that of a Casimir energy. Comparing coefficients, one finds that the $1/L$
term we found in \aaa\ corresponds to a L{\"u}scher term in 2+1 dimensions.
It is hard to believe that this is a
coincidence given that the AdQCD$_2$ can be thought as the dimensionally
reduced theory from the 2+1 high-Temperature QCD. We thus have the
rather intriguing result that the 2 dimensional string remembers its 3
dimensional origin via the presence of the adjoint scalar field.

We also again want to stress that we know of no direct field theoretical way to
arrive at our results.

Finally, we emphasize that in spite of all approximations, we have a
very definite prediction for the heavy quark potential as a function of
the renormalized mass. A lattice Monte Carlo simulation of the theory
should tell us if we are on the right track or not.

\ack{We would like to thank M. Douglas, I.R. Klebanov, J. Greensite,
J. Minahan, H. Nielsen, H. Rubinstein and I. Zahed
for discussions. We especially thank A. P. Polychronakos for many
enlightening conversations and a critical reading of the manuscript.}

\bigskip
\noi
{\bf Appendix.  Expectation values of Wilson loops }

In this appendix we discuss the method used in section 2 for computing
the Wilson loop averages and taking the large $N$ limit.
We also give a sample calculation.
For the non-intersecting loops it is illuminating to use the Hamiltonian
formalism of ref. [\poly].
In this formalism one lets time run in one direction and the
Wilson loops are equal-time loops that wrap around the spacial direction.
The variables of the theory are the Wilson loops (W) while
the states are given by the characters $\X _R$ of the group in the
specific representation. The propagation of
a state in the representation $R$ through an area $S$ is achieved by the
operator $ e ^ { -{ g_0 ^ 2  \over N } S C _R}$, where $C_R $ stands for
the quadratic casimir of the representation $R$.

We first consider the configuration of a single adjoint loop inside a
fundamental.
We denote the quantity to be computed by $\left< W _F W _ A \right>$,
where the subindices
refer to the representations of the loop, fundamental or adjoint
correspondingly.
Because of the infinite limit of the outside area, the only state which
propagates from $t = - \infty $ until the point
of the fundamental loop insertion is the trivial one (vacuum).
Then
$$
 \langle W_F W_A \rangle =\int \, DW \, \X _ F \, e ^ {- { g_0 ^ 2 \over N }
S _ 1 C _F } \X _ A G ( t _A - t _F) \sum _ {\tilde R _ j } \X _
{\tilde R _ j } d _ { \tilde R _ j }
\eqn\fa$$
where $\X _ F $ and $ \X _A$ denote the insertions of the fundamental and
adjoint loops correspondingly, $ G (t _A -t_F)$ stands for the propagator
from the point of the
fundamental loop insertion $ t _F$ until the point of the adjoint loop
$t _A$ and
$ \sum \X _ {\tilde R _ j}$ is a complete set of states inserted at the time
$t = \infty$. The factor $d_{ \tilde R _j} $ denotes the dimension of the
representation $\tilde R _ j$ and it is inserted for each representation
which reaches the ``north'' pole.

If $ F \otimes A = R_1 \oplus R _2 \oplus F$, then for the
characters of the group holds
$$
\X _ F \cdot \X _ A = \X _ { F \otimes A} = \X _ {R _ 1} + \X _ {R _ 2 }
+ \X _ F
\eqn\c$$
Inserting the last expression in \fa\ and using the orthogonality
condition
$$
\int DW \X _ {R _ i} (W) \X _ {\tilde R _ j} (W) = \delta _ { R _ i ,
\tilde R _ j }
\eqn\oc$$
we obtain
$$
\left< W _ F W _ A \right>  =
e ^ {- {g_0^2 \over N}(S_F -S_1) C_F }
\left( d _ F e ^ { - {g_0^2 \over N}  S_1 C _F}
+ d _ {R_1} e^{ - { g_0 ^ 2 \over N }  S _1 C_{R_1}}
+ d _ {R_2} e^{ - { g_0 ^ 2 \over N } S_1 C_{R_2}} \right)
\eqn\rf$$
where $S_F $ and $S_1$ are the areas of the
fundamental and adjoint loops respectively.

The Hamiltonian approach used in obtaining the last result is
not suitable for intersecting or self-intersecting loops. The reason
is that at the point of intersection
the breaking of the space results to correlations between the two pieces
and as a result the simple Hamiltonian rules are not applicable.
Alternatively, one could use path-integral formulation in order to obtain
the results for these more complicated topologies. Some explicit expression
for (self)intersecting loops,
obtained by this method, are given in section 3 in the text.
Notice at this point that all results for (self)intersecting loops
are for adjoint loops
without the presence of the external fundamental loop. (The calculation
of expectation values of (self)intersecting Wilson loops inside an external
loop is a very difficult problem and has not been solved explicitly.)

In order to study the large $N$ limit of our theory
we use the following group theoretical arguments for the dimension
and the quadratic casimir of the product of two representations:
assume that $\tilde R _i$ are the irreducible components contained
in the decomposition of the product of the representations $R_1$ and
$R _2$, that is,
$ R_1 \otimes R_2 =\sum _ i \tilde R _ i $, where $i =1, ... \nu$.
Then, in the large $N$ limit there exist
some representations $\tilde R _l$, where $l=1, ..\mu $ and $\mu < \nu $,
such that to leading
order in $N$ the dimension of the product of the representations are
given by suming the dimensions of these leading components, namely,
$\, \sum _ l d  {\tilde R _ l} = d {R _1} \cdot d {R_2} \equiv
d (R _ 1 \otimes R _ 2)$.
Moreover, the casimirs of these leading representations are all
equal to the sum of the casimirs of the original ones, again, to
leading order in $N$.
Therefore for a decomposition of two such representations the $\tilde
R _l$ are the ones that will contribute in the large $N$ limit.

Using these group theoretical rules in \rf\ we obtain the following large $N$
expression
$$
\langle W _F W_A \rangle = d _F e ^ { - {g_0 ^2 \over N} S _F C_F }
d _A e ^ {- {g_0 ^2 \over N } S_1 C_A}
\eqn\ln$$

\refout
\vfill\eject

\bigskip\noi
{\bf Figure Captions}
\medskip\noi
Fig. 1a. Non-intersecting adjoint loops inside a fundamental.
\medskip\noi
Fig. 1b. Overlapping, non-intersecting adjoint loops inside a
fundamental.
\medskip\noi
Fig. 2a. Intersecting adjoint loops.
\medskip\noi
Fig. 2b. A self-intersecting adjoint loop.
\medskip\noi
Fig. 2c. A self-intersecting adjoint loop with overlapping areas.
\medskip\noi
Fig. 3. A configuration with a section, of area $S$,  of a large adjoint loop
outside of the fundamental, is weighted by the factor
$e^{-S { \sigma }_A} $, and therefore suppressed for large $S$.
\vfill
\eject
\end